\documentclass{article}
\usepackage{graphicx}

\usepackage{amsmath}
\usepackage{booktabs}
\PassOptionsToPackage{hyphens}{url}
\usepackage{hyperref}
\usepackage{graphicx}
\usepackage{xparse}
\usepackage{ragged2e}

\newlength{\cardheight}
\newlength{\widecardheight}
\usepackage{array}
\usepackage{tabularray}
\ExplSyntaxOn
\cs_if_exist:NF \tbl_save_outer_table_cols: { \cs_new_protected:Npn \tbl_save_outer_table_cols: {} }
\ExplSyntaxOff
\usepackage{multirow}
\usepackage{booktabs}
\usepackage[authoryear,round]{natbib}
\usepackage{xcolor}
\usepackage{comment}
\usepackage{fontspec}
\usepackage{listings}
\usepackage{polyglossia}

\usepackage{subcaption}
\usepackage{float}

\setdefaultlanguage{english}
\setotherlanguages{arabic,hebrew}

\newcolumntype{C}[1]{>{\centering\arraybackslash}p{#1}}
\title{Israel-Hamas War on X: A Case Study of Coordinated Campaigns and Information Integrity}

\author{}
\date{}

\begin{document}

\maketitle
\vspace{-2.5em}
\begin{center}
  Tu\u{g}rulcan Elmas$^{1,3}$,\;
  Filipi Nascimento Silva$^{2,3}$,\;
  Manita Pote$^{3}$,\;
  Priyanka Dey$^{4}$\\[0.3em]
  Keng-Chi Chang$^{5}$,\;
  Jinyi Ye$^{4}$,\;
  Luca Luceri$^{4}$,\;
  Cody Buntain$^{6}$,\;
  Emilio Ferrara$^{4}$\\[0.3em]
  Alessandro Flammini$^{3}$,\;
  Filippo Menczer$^{3}$\\[0.8em]
  {\footnotesize
  $^{1}$University of Edinburgh\quad
  $^{2}$Northwestern University\quad
  $^{3}$Indiana University Bloomington\\[0.25em]
  $^{4}$University of Southern California\quad
  $^{5}$Dartmouth College\quad
  $^{6}$University of Maryland}
\end{center}
\vspace{1em}

\begin{abstract}
Coordinated campaigns on social media play a critical role in shaping crisis information environments, particularly during the onset of conflicts when uncertainty is high and verified information is scarce. We study the interplay between coordinated campaigns and information integrity through a case study of the 2023 Israel-Hamas War on Twitter (X). We analyze 4.5~million tweets and employ established coordination detection methods to identify 11 coordinated groups involving 541 accounts. We characterize these groups through a multimodal analysis that includes topics, account amplification, toxicity, emotional tone, visual themes, and misleading claims. Our analysis reveal that coordinated campaigns rely predominantly on low-complexity tactics, such as retweet amplification and copy-paste diffusion, and promote distinct narratives consistent with a fragmented manipulation landscape, without centralized control. Widely amplified misleading claims concentrate within just three of the identified coordinated groups; the remaining groups primarily engage in advocacy, religious solidarity, or humanitarian mobilization. Claim-level integrity, toxicity, and emotional signals are mutually uncorrelated: no single behavioral signal is a reliable proxy for the others. Targeting the most prolific spreaders of misleading content for moderation would be effective in reducing such content. However, targeting prolific amplifiers in general would not achieve the same mitigation effect. These findings suggest that evaluating coordination structures jointly with their specific content footprints is needed to effectively prioritize moderation interventions.
\end{abstract}

\section{Introduction}

Modern armed conflicts unfold in both the physical realm and in the information environment, codified as ``Operations in the Information Environment'' in US military doctrine.
In these operations, narratives, images, and emotions shape global perceptions in real time, extending traditional propaganda strategies to win hearts and minds~\citep{10.1162/152039702753649656}.
The Syrian Civil War marked one of the first large-scale conflicts in which social media became central to documenting violence, mobilizing international sympathy, and disseminating propaganda~\citep{ayad2019baghdadi,walk2025social}.
Russia's 2022 invasion of Ukraine demonstrated an even more sophisticated digital information environment, characterized by coordinated state messaging, rapid narrative framing, and multilingual misinformation campaigns~\citep{boyte2017analysis, Dukach2025, pierri2023moderation, geissler2023russian, lai2024multilingual}.
The 2023 Israel-Hamas War further exemplified this evolving information battleground. As violence erupted on October~7, 2023, Twitter (now X) became a central arena for contesting narratives, with adversaries deploying fabricated videos~\citep{Brooking2024Distortion}, misleading images~\citep{ArabCenterDC2023Gaza}, and coordinated posts~\citep{AlJazeera2025Spinning, Hendrix2024Evaluating} from the earliest hours.
Susceptibility to misinformation and propaganda in this war context depends on the cognitive, moral, and socio-political characteristics of targeted individuals~\citep{doi:10.1111/pops.70127}.

Prior work has found that coordinated online activity is structurally heterogeneous \citep{luceri2024unmasking,ng2022online}: harmful narratives tend to concentrate in specific clusters rather than spreading uniformly, and a small fraction of accounts can drive the bulk of amplification~\citep{alieva2024propaganda, zia2023analysis,pierri2022propaganda, nogara2022disinformation,deverna2024identifying}.
Existing research addresses either coordination signals or content integrity in isolation. For instance, coordination-detection studies~\citep{geissler2023russian,pierri2023moderation,pacheco2021uncovering,vishnuprasad2024tracking,nizzoli2021coordinated} identify groups of accounts acting in concert but do not systematically assess which groups actually amplify misleading claims.
Conversely, content-focused analyses of conflict misinformation~\citep{lai2024multilingual,shahi2025too} flag harmful narratives but lack the network resolution to attribute integrity risk to specific coordinated actors.
Detecting coordination and evaluating its information-integrity implications during conflicts thus remain largely disconnected tasks.

The present study bridges this gap through an analysis of coordinated activity on Twitter\footnote{The platform was renamed X after the study period, so we henceforth refer to it as Twitter.} during the onset of 2023 Israel-Hamas War. We focus on the period between September~1 to December~15, 2023, a period with high uncertainty, scarce verified information, and intense global attention: conditions under which coordinated campaigns may be especially consequential. We apply established coordination-detection techniques \citep{pacheco2021uncovering} to identify 11 groups involving 541 accounts that were likely coordinated rather than independent in their posting patterns. We characterize each group across narratives, claim-level information integrity, toxicity, emotions, retweeted accounts, tweet types, and image use.

We address the following research question: \textit{What were the structural, behavioral, and narrative characteristics of coordinated information campaigns during the onset of the Israel-Hamas War on Twitter, and to what extent did they amplify misleading claims?}
Our study makes four contributions:
\begin{enumerate}
    \item We provide the first large-scale, multimodal characterization of coordinated campaigns during the 2023 Israel-Hamas War on Twitter, revealing multiple coordinated groups driven by two low-complexity tactics: co-retweet amplification and copy-paste diffusion.
    The coordinated groups differ substantially in content profiles and behavioral signatures. Each campaign is organized around a small set of amplified accounts that generally receive little engagement from non-coordinated accounts, suggesting that these campaigns fail to deeply penetrate organic user networks.

    \item We assess the integrity of the claims found in 191 highly amplified coordinated posts, identifying 12 misleading posts. These posts are concentrated in three of the coordinated groups, indicating that information-integrity risks are not uniformly distributed across coordinated groups.

    \item We show that claim-level integrity, toxicity, and emotion signals are mutually uncorrelated: no single content-based signal is a reliable proxy for the others.

    \item We demonstrate that removing the most prolific spreaders of misleading content is an effective moderation method while also exhibiting diminishing returns. By contrast, targeting prolific amplifiers of all widely retweeted content without prior knowledge of which posts are misleading does not achieve the same mitigation effect.
\end{enumerate}

Although our empirical focus is the Israel-Hamas War on a single platform, the proposed group-level integrity-risk framework is applicable to other settings where networked coordination and content integrity are at play. Our results suggest that integrating coordination structure with content integrity signals offers a more precise basis for moderation than treating all coordinated activity as uniformly harmful.

\section{Related Work}

Coordination on social media has emerged as a key tactic in online campaigns, enabling a single entity to orchestrate the activities of multiple accounts and amplify specific narratives. Coordinated accounts typically manipulate information visibility and create the false appearance of widespread support or consensus. Below, we survey detection methods and their applications, report on findings about misinformation in conflicts, and summarize prior work on the 2023 Israel-Hamas War.

\subsection{Coordination Detection}

Coordination detection leverages digital traces left behind by coordinated accounts: identical text they share, posting times, and collectively amplifying hashtags, accounts, or URLs.
\citet{pacheco2021uncovering} formalize several of these signals---synchronized posting, co-retweeting, sharing indentical images or hashtag sequences---into a general framework, as applied to political campaigns.
\citet{luceri2024unmasking,federico2025cross,anand2025density} employ network properties, such as density and centrality, in similarity networks for detection.
Other signals include coordinated-reply attacks targeting influential figures \citep{pote2025coordinated}, ephemeral posts and likes designed to manipulate ranking algorithms while deleting traces \citep{elmas2021ephemeral,Torres2022deletions}, mutual follows and trains \citep{elmas2024teamfollowback,torres2022manufacture}, promotion of suspicious web domains and mock news sites \citep{minici2024uncovering}, conspiracy-theory amplification \citep{vishnuprasad2024tracking}, and using generative AI to generate human-like content \citep{yang2023anatomy,luceri2025coordinated}.
A recurrent finding is that effective coordination need not be technically sophisticated: text-template campaigns (a.k.a. ``copy-pastas'') linked to astroturfing behavior that simulates grassroots consensus \citep{vishnuprasad2024tracking,keller2020political} and concentrated retweeting \citep{elmas2022characterizing,astrofurfSchoch} remain among the simplest yet most persistent amplification tactics.
Despite concerns around the growing sophistication of influence campaigns, our analysis reveals similarly unsophisticated instances of coordination that also rely on concentrated retweeting and copy-pastas.

Coordination detection has been applied to uncover campaigns in many domains such as elections \citep{nizzoli2021coordinated,ng2022online}, pump-and-dump cryptocurrency schemes \citep{nizzoli2020,mirtaheri2021crypto,pacheco2021uncovering}, state-sponsored information operations \citep{sio,cima2024coordinated}, fake Twitter trends \citep{gopalakrishnan2025large}, cross-platform link-sharing strategies \citep{federico2025cross}, and multimedia campaigns on TikTok \citep{luceri2025coordinated}. We contribute to this line of research by analyzing coordination during the onset of a war and pairing coordination detection with systematic integrity assessment. We survey the conflict-specific coordination literature in the next subsection.

\subsection{Coordination and Misinformation in Conflicts}

High uncertainty, scarce verified information, and global attention make the public more vulnerable to manipulation campaigns during armed conflicts. The 2022 Russia-Ukraine War exemplified this phenomenon.
\citet{pierri2023moderation} find that coordinated account creation peaked on the invasion's first day (February~24, 2022), with newly created accounts disproportionately posting replies, spam, and hate speech rather than original content.
\citet{geissler2023russian} estimate that roughly one in five pro-Russian content spreaders were bots, many created at the onset of the invasion and located in countries that abstained from the UN vote condemning the invasion.
\citet{pierri2022propaganda} report that amplification during the war was highly concentrated, studying 19.5 million Facebook posts and finding that the top 15 accounts comprised 60--80\% of all propaganda reshares.
\citet{lai2024multilingual} identify five dominant families of false narratives circulating in six languages during the invasion's first two weeks, with different language communities propagating different subsets.

Two findings from this literature are directly relevant to the present study. First, harmful content does not spread uniformly among coordinated actors: toxicity concentrates in specific ideologically aligned communities \citep{zia2023analysis}. Second, dominant propaganda frames such as the ``fascist/nazi'' narrative in the Russia-Ukraine War are concentrated in a few communities \citep{alieva2024propaganda}.
Our study corroborates these findings by showing that toxicity, misinformation, and elevated emotions are concentrated in select communities.
Such a structural heterogeneity motivates our decision to assess integrity risk at the coordinated community level rather than treating all detected coordination as uniformly harmful.
\citet{kim2024toxic} further show that hate speakers on Twitter are exposed to disproportionately more low-credibility news sources, suggesting a potential synergy between toxic language and misinformation. We do not find such a relationship in the Gaza War case.

\subsection{The 2023 Israel-Hamas War Online}

A growing body of work has begun to examine online discourse surrounding the 2023 Israel-Hamas War.
The DFRLab identified a network of over 25 coordinated Twitter accounts \citep{dfrlab2023southasian}.
Many of them self-reported to be in India.
They spread copy-pasted disinformation during the conflict's first eleven days, switching allegiances between pro-Israel and pro-Palestinian positions.
This verbatim-text coordination is analogous to that of some communities we detect here.
\citet{antonakaki2025israel} analyze topics and sentiment using datasets from Telegram, Reddit, and Twitter but similarly stop short of examining coordinated activity and network-level manipulation due to the scale of their dataset (2,000 tweets).
\citet{ng2025prominent} examine posts across Twitter, Reddit, and TikTok during the conflict's first four months, identifying major engagement spikes around key events (the Al-Ahli hospital explosion, the Al-Shifa hospital raid) through peaks in data volume and persistent negative sentiment.
These studies primarily describe discourse-level trends or platform-wide sentiment rather than systematically studying coordinated manipulation and misleading content. Our work addresses this gap by comprehensively integrating coordination-network structure analysis, claim-level integrity assessment, and multimodal behavioral characterization (toxicity, emotion, visual themes) on a large-scale dataset.

\section{Data}

For the present analysis, we leverage a dataset provided as part of a challenge problem in the DARPA-sponsored Influence Campaign Awareness and Sensemaking (INCAS) program.\footnote{\href{https://www.darpa.mil/research/programs/influence-campaign-awareness-and-sensemaking}{www.darpa.mil/research/programs/influence-campaign-awareness-and-sensemaking}}
This dataset consists of tweets related to the Israel-Hamas War, predominantly posted between October~7 and December~15, 2023 (less than 0.5\% of the tweets were posted in the preceding month).

Tweets were selected based on inclusion of at least one term from each of two sets of keywords: one related to Palestine, Palestinian, Gaza, West Bank, or Jerusalem, including Arabic and Hebrew variants and transliterations; and one related to Israel or Israeli, including Arabic and Hebrew variants and transliterations.
The dataset excludes tweets with explicit, pornographic, racist, or otherwise derogatory keywords, as well as those referring to major sports and US politics.

Table~\ref{tab:dataset} reports the distribution of tweet types and the numbers of users, images, and original tweets with images. The dataset comprises 4.5 million tweets by 1.5 million accounts. Retweets dominate, indicating amplification behavior commonly observed in coordinated influence campaigns. 93.5\% of the tweets are in English, 5.4\% Arabic, 0.5\% Indonesian, 0.04\% Hindi, 0.03\% Hebrew, 0.01\% Persian, and the remainder in other languages.

\begin{table}
\centering
\caption{Summary of dataset. The image count is based on unique image URLs appearing in any tweet.}
\begin{tabular}{l r r}
    \hline
    \textbf{Description} & \textbf{Count} & \textbf{Percentage (\%)} \\
    \hline
    Original tweet & 216,144 & 4.7 \\
    Retweet & 4,012,031 & 87.8 \\
    Quote & 61,873 & 1.4 \\
    Replies & 280,585 & 6.1 \\
    \hline
    Total tweets & 4,570,633 & 100.0 \\
    \hline
    Users & 1,521,433 & -- \\
    Images & 268,434 & -- \\
    Original tweets with images & 86,487 & 40.0 \\
    \hline
\end{tabular}
\label{tab:dataset}
\end{table}

\section{Coordination Detection}
\label{sec:coordination_detection}

\begin{figure}
    \centering
    \includegraphics[width=\linewidth]{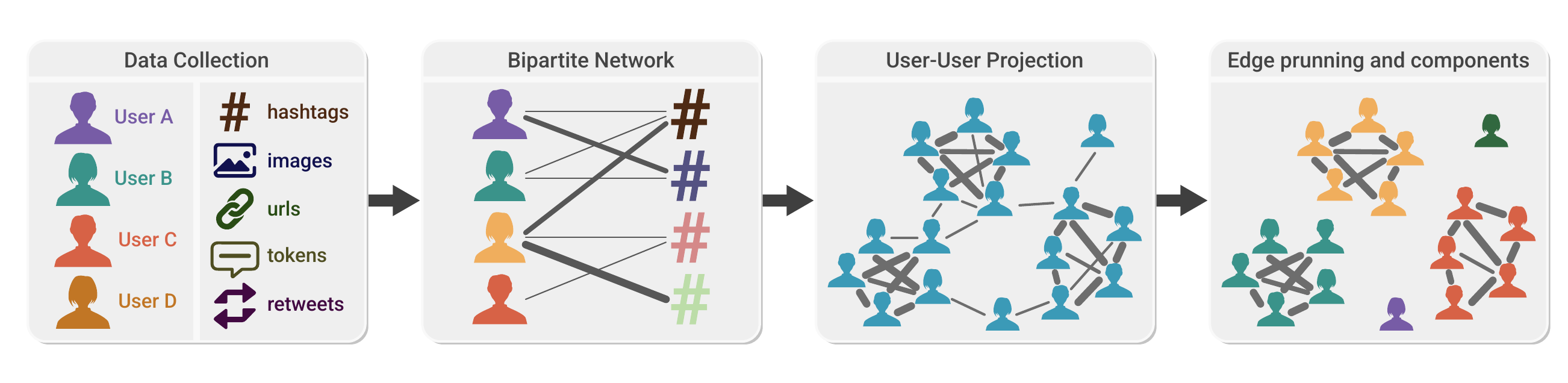}
    \caption{Overview of the coordination detection pipeline.}
    \label{fig:overview}
\end{figure}

To identify coordinated groups of accounts, we analyze co-sharing of text-based content (hashtags, URLs, similar text, and retweeted accounts) and image-based content. We apply the framework proposed by \citet{pacheco2021uncovering}, which looks for patterns in which multiple accounts share the same or highly similar items, referred to as \textit{indicators}. Such overlap can signal coordinated activity when statistically unlikely. Figure~\ref{fig:overview} shows an overview of our approach, which we describe in detail next.

We first build a collection of bipartite graphs that connect user nodes and indicator nodes, where each graph corresponds to one of five indicator types. An edge connects a user to an indicator if the user engages with that item, with edge weight equal to the engagement frequency.
We first filter out infrequent indicator nodes that are used by fewer than five distinct accounts. We then filter out low-activity accounts that are connected to less than five distinct indicator nodes within each bipartite graph.

\begin{figure}
\begin{center}
\includegraphics[width=0.7\textwidth]{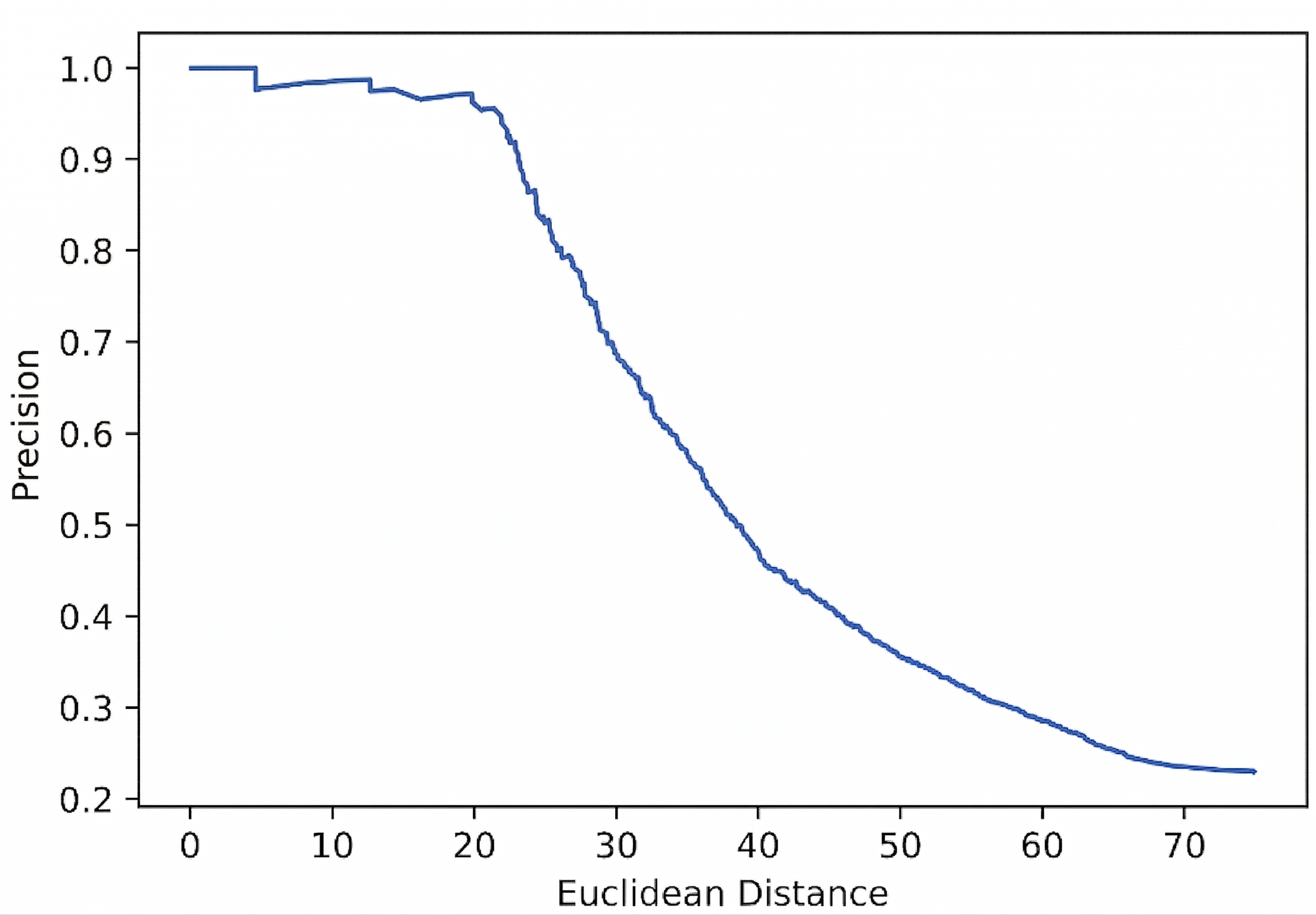}
\caption{Precision as a function of Euclidean distance threshold. Image pairs with distance below 10 in the BLIP-2 embedding space are highly likely to be duplicates.}
\label{fig:dup_rate}
\end{center}
\end{figure}

Each indicator type specifies what constitutes a shared item, applies to one or more types of post, and may require specific preprocessing:

\begin{enumerate}

\item \textbf{Retweeted account.} The retweet indicator is applied to retweets. Each retweeted user account constitutes a single indicator node, and an edge is added between a retweeting user and the retweeted account, with edge weight equal to the number of times the same user retweeted the same account.

\item \textbf{Hashtag.} The hashtag indicator is applied to non-retweet posts (original tweets, quotes, and replies). Each hashtag used in a post becomes an indicator node, and an edge is added between a user and a hashtag mentioned by that user, with weight equal to the number of posts by the same user including the same hashtag.

\item \textbf{URL.} The URL indicator is applied to non-retweet posts (original tweets, quotes, and replies). Before indexing, tracking and analytics parameters are stripped from each URL. Each unique cleaned URL becomes an indicator node, and an edge between a user and a URL indicates how many times that user shared that link.

\item \textbf{Token.} The token indicator is applied to non-retweet posts (original tweets, quotes, and replies).

Each post undergoes a cleaning stage in which URLs, @mentions, digits, HTML tags, emojis, hashtags, punctuation, and leading ``RT'' tokens are removed. The cleaned text is then tokenized and normalized through lemmatization and lowercasing. Tokens with fewer than three characters and English stopwords are further removed.
Each unique remaining token constitutes an indicator node, and an edge is added between a user and a token for every occurrence of that token in the user's posts, with edge weight corresponding to frequency.

\item \textbf{Image.} The image indicator requires resolving near-duplicates: the same image frequently recirculates in slightly modified forms (cropped, recolored, or watermarked), so that exact-match grouping would artificially fragment what is effectively the same shared content. To group near-duplicate images into a single indicator node, we employ an approach by \citet{Joshi_Buntain_2024}. We embed images using BLIP-2 \citep{pmlr-v202-li23q} and compute pairwise Euclidean distance among these embeddings.
We then establish a threshold such that an image pair with distance below this threshold is treated as a duplicate or near-duplicate pair.
We determine the threshold via manual qualitative assessment. We first randomly sample 618 image pairs from our collection and assign each pair to one of four classes: duplicate, near-duplicate (i.e., the same image with minor modifications), different, or unknown.
We label 106 image pairs as exact duplicates, 36 as near-duplicates, and 476 as different.
To find a reasonable distance threshold, we plot the threshold versus the precision (the number of duplicate/near-duplicate labels over the number of image pairs with distance below the threshold) in Figure~\ref{fig:dup_rate}. We manually select a conservative threshold of 10, yielding 0.98 precision.
Using this threshold, we get 0.64 recall of duplicates and 0.48 of near-duplicates/duplicates.
After merging duplicates and near-duplicates, each image is treated as an indicator node, and an edge is added between a user and an image shared by that user, with edge weight corresponding to the number of times the same image is shared by the same user.

\end{enumerate}

Table~\ref{tab:summary_stats} summarizes the scale of each indicator's bipartite graph. The retweet indicator covers by far the most users, consistent with the dominance of retweet activity in the dataset.

\begin{table}
    \small
    \centering
    \caption{Summary statistics of bipartite graphs underlying each coordination indicator. I/U and U/I columns report the average number of indicators per user and vice versa.
    Indicators with higher U/I values are shared by many users, producing stronger coordination signals.}
    \label{tab:summary_stats}
    \begin{tabular}{lrrccc}
    \hline
    \textbf{Indicator} &
    \textbf{Users} &
    \textbf{Indicators} &
    \textbf{I/U} &
    \textbf{U/I} &
    \textbf{Avg. Weight} \\
    \hline
    Hashtags & 6,643 & 16,389 & 2.47 & 0.41 & 4.51 \\
    URLs &       227 & 500    & 2.20 & 0.45 & 1.36 \\
    Retweets & 159,120 & 43,807 & 0.28 & 3.63 & 1.00 \\
    Tokens & 49,339 & 19,025 & 0.39 & 2.59 & 1.75 \\
    Images & 36,455 & 72,856 & 2.00 & 0.50 & 1.02 \\
    \hline
    \end{tabular}
\end{table}

We project the bipartite graph of each indicator into a user-user similarity network.
In this co-indicator network, the weight of an edge between two users is computed as the cosine similarity between the TF-IDF vectors constructed from the indicators associated with each user.
The idea is that two users with high co-retweet similarity repeatedly amplify the same set of accounts.
Similarly, users with high co-hashtags similarity deploy the same set of hashtags, users with high co-URL similarity link to the same resources, and users with high co-token similarity use the same lexical items, capturing similarity in language usage rather than exact text duplication.
Finally, co-image similarity identifies accounts that repeatedly share the same or highly similar images.

To prune incidental connections, we wish to retain only the most similar pairs of users in each projected co-indicator network.
We therefore rank all similarity scores and keep only the top $10^{-5}$ fraction of highest-weight edges, considering all nodes with at least one non-zero similarity across all the indicators (49,333 nodes).
This thresholding step ensures that the analysis focuses on the most meaningful coordination signals while filtering out noise arising from coincidental similarities.
The resulting networks contain clusters of users connected by strong coordination ties.

Finally, we merge the five pruned co-indicator networks into a single coordination network, similar to the approach proposed by \citet{luceri2024unmasking}.
The merge is performed as a union of edges: if any coordination indicator (retweet, hashtag, URL, token, or image) links a pair of users, that connection is included in the combined network.
We treat the resulting network as unweighted, since our focus is on whether a strong coordination tie exists rather than the exact magnitude of the original similarity score.

In the merged coordination network, connected components represent groups of users linked through one or more coordination indicators.

While many components are small (e.g., dyads or triads), our analysis focuses on the most substantial communities, which are more likely to reflect potential coordinated campaigns or factions. We therefore exclude components with fewer than six user nodes. Among the remaining 11 groups, we observe a natural split in structure: small components (6--11 nodes) tend to form cliques or near-cliques, while larger components (16 nodes and above) exhibit sparser clustering. Together, these 11 components encompass 542 accounts --- 1.1\% of the 49,333 users with non-zero coordination similarity.

The components in this collection include only edges bases on co-retweet and co-token indicators.
Co-hashtag, co-URL, and co-image indicators produce no user pairs surviving the edge-pruning process described above.
Further, we did not observe mixed coordination, i.e., the same pair of users coordinating through multiple indicators.
This finding aligns with prior research showing that influence campaigns frequently employ multiple, minimally overlapping subnetworks that use different strategies, presumably to evade detection \citep{luceri2024unmasking,ng2022online,luceri2025coordinated}.

We analyze each component manually based on its most salient amplification behavior.
For co-retweet components, we use their most prevalent retweet target(s).
For co-token components, we observe repeated text templates posted across accounts (``copy-pastas''), and we name these components after those repeated templates \cite{vishnuprasad2024tracking}. Figure~\ref{fig:components} provides an overview of the components. We reference components using a numeric identifier and compact alias (e.g., Component~1\,-\,Khamenei) throughout the paper, and provide full component names in Section~\ref{sec:component_narratives}.

\section{Characterization}
\label{sec:characterization}

\NewDocumentCommand{\compcard}{m m m m m m m}{%
  \begin{minipage}[t][\cardheight][t]{0.325\textwidth}
    \vspace{0pt}%
    \begin{minipage}[t]{0.25\linewidth}
      \vspace{0pt}%
      \includegraphics[width=\linewidth]{#1}
    \end{minipage}\hfill%
    \begin{minipage}[t]{0.72\linewidth}
      \vspace{0pt}%
      \RaggedRight
      \textbf{#2}\par
      \vspace{0.28em}
      {\scriptsize #3 users $\cdot$ #7 tweets $\cdot$ \textit{#4}\par}
      \vspace{0.35em}
      {\tiny
      \textbf{Orig:} #5\par
      \vspace{0.18em}
      \textbf{RT:} #6\par}
    \end{minipage}%
  \end{minipage}%
}

\NewDocumentCommand{\compcardwide}{m m m m m m m}{%
  \begin{minipage}[t][\widecardheight][t]{0.495\textwidth}
    \vspace{0pt}%
    \begin{minipage}[t]{0.28\linewidth}
      \vspace{0pt}%
      \includegraphics[width=\linewidth]{#1}
    \end{minipage}\hfill%
    \begin{minipage}[t]{0.68\linewidth}
      \vspace{0pt}%
      \RaggedRight
      \textbf{#2}\par
      \vspace{0.28em}
      {\scriptsize #3 users $\cdot$ #7 tweets $\cdot$ \textit{#4}\par}
      \vspace{0.35em}
      {\tiny
      \textbf{Orig:} #5\par
      \vspace{0.18em}
      \textbf{RT:} #6\par}
    \end{minipage}%
  \end{minipage}%
}

\begin{figure}
\centering
\small

\compcardwide
  {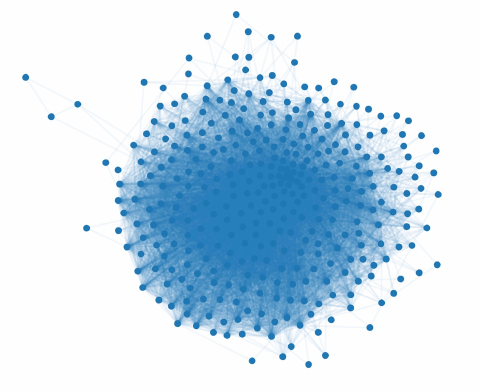}
  {1\,-\,Khamenei}
  {344}
  {coretweet}
  {israel zionist, terrorist state, free palestin}
  {storm, expect storm, storm nations, regime duty, stand zionist}
  {131,440}
\hfill
\compcardwide
  {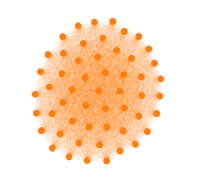}
  {2\,-\,FCNL}
  {53}
  {cotoken}
  {war civilian always, always pay high, grow lawmaker}
  {gaza, israel, kill, weapon, genocide}
  {33,596}

\vspace{1.0em}

\compcard
  {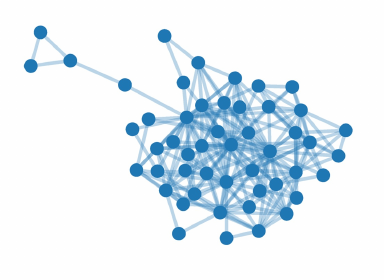}
  {3\,-\,SoftWarNews}
  {50}
  {coretweet}
  {bomb civilization, israel bomb, palestine}
  {palestina, yang, ini, anak, akan}
  {13,513}
\hfill
\compcard
  {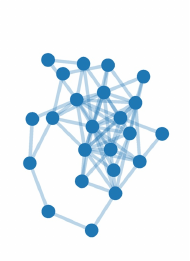}
  {4\,-\,Ahmadiyya}
  {23}
  {coretweet}
  {globe leader, civilian muslims, true peace justice}
  {muslim nation protect, palestinians watch, full episode}
  {16,413}
\hfill
\compcard
  {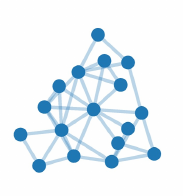}
  {5\,-\,AntiIran}
  {17}
  {coretweet}
  {europe day, ago knife, youth europe, attack}
  {gaza, irgc, israel, calculus, nujaba}
  {3,392}

\vspace{1.0em}

\compcard
  {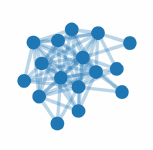}
  {6\,-\,HuT}
  {16}
  {coretweet}
  {bbc promote bias, propaganda pro israeli}
  {twitter storm, bless land, land aqsa, gaza}
  {6,621}
\hfill
\compcard
  {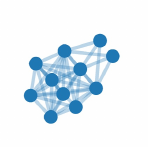}
  {7\,-\,Tanzeem}
  {11}
  {coretweet}
  {(no original tweets)}
  {Islamic greetings (Arabic/Urdu)}
  {2,163}
\hfill
\compcard
  {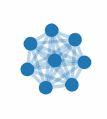}
  {8\,-\,DFRAC}
  {8}
  {coretweet}
  {(no original tweets)}
  {gaza, israel, fact check (Hindi)}
  {940}

\vspace{1.0em}

\compcard
  {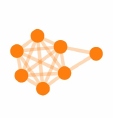}
  {9\,-\,Starlink}
  {7}
  {cotoken}
  {internet via starlink, help provide internet}
  {gaza, israel, kill, weapon, genocide}
  {4,743}
\hfill
\compcard
  {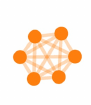}
  {10\,-\,Hospital}
  {6}
  {cotoken}
  {baptist hospital bad day, war propagandist}
  {gaza, israel, kill, weapon, hamas}
  {1,351}
\hfill
\compcard
  {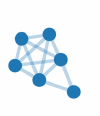}
  {11\,-\,USSec}
  {6}
  {coretweet}
  {(no original tweets)}
  {gaza, israel, macgregor threat, homeland iran}
  {1,955}

\vspace{0.8em}

\caption{Detected coordinated components in the merged coordination network. Each panel shows the component's network, alias, user count, total tweet count, coordination type (also encoded in edge colors), and distinguishing keywords from original tweets and retweets.}
\label{fig:components}

\end{figure}

We next characterize each component by identifying the key features that distinguish it from other components and the rest of the dataset. We examine each component along multiple dimensions: (i)~dominant narratives and topics its members promote through keywords and hashtags, (ii)~claim-level information integrity in the most salient amplified posts, (iii)~distribution of tweet types, (iv)~most retweeted accounts, (v)~prevalence of toxic or abusive language, (vi)~emotional tone of the shared content, and (vii)~visual themes in shared images. We describe the methodology for each dimension and provide key results below. In Section~\ref{sec:component_narratives}, we synthesize these dimensions into per-component narrative profiles and highlight components that amplify misleading, false, or contested claims.

\subsection{Topical Analysis}

To summarize the dominant narratives promoted by each coordinated component, we identify the hashtags and keywords that most clearly distinguish it from the full dataset. For each connected component, we aggregate all tweets authored by its members. We then find the words that occur disproportionately frequently within the component by calculating the log-odds ratios \citep{monroe2008fightin}.
This analysis provides a ranked list of component-specific terms indicative of each group's topical focus. We use these topics in Section~\ref{sec:component_narratives}.

\subsection{Information Integrity}

To assess information integrity, we annotate posts that were retweeted by at least five accounts within a component (191 posts total). We define a post as \emph{misleading} if the claim it makes is verifiably false, factually contested, or presented in a way that could mislead the reader (for instance, through decontextualization, selective omission, or attribution to fabricated sources). We provided this definition and the 191 posts to GPT 5.2 to classify if each post includes a misleading claim or not. One author read and manually verified the 12 posts GPT 5.2 classified as misleading by cross-checking them with external fact-checking evidence. The same annotator manually grouped these misleading posts into three claim families as they refer to three separate events. We describe the three labeled claim families below; per-component engagement with each claim is presented in Section~\ref{sec:component_narratives}.

\begin{itemize}
    \item \textbf{Claim~C1 - Hospital attribution and ``deleted evidence'' (6 posts):} On October~17, 2023, an explosion in and around the Al-Ahli Arab (Baptist) Hospital in Gaza City killed a large number of civilians during intense early-war chaos. In the first hours after the event, actors across the information ecosystem made conflicting attribution claims. The \textit{hospital-strike} variant consists of posts that state or strongly imply that Israel directly struck the hospital site, while the \textit{deleted evidence} variant treats deleted posts by @HananyaNaftali (Israel's digital spokesperson at the time) and @Israel (the official state account) as conclusive proof of responsibility. This framing goes beyond what the deletions alone establish, and attribution remains contested across independent investigations~\citep{politifact_naftali_2023,hrw_ahli_2023,forensic_architecture_ahli_2024}. Component~1-Khamenei is the primary amplifier: the hospital-strike variant was retweeted by 143~accounts within the component and the deleted evidence variant by 99~accounts. Component~10-Hospital coordinates around the same narrative as its central copy-pasta (see Section~\ref{sec:component_narratives}).

    \item \textbf{Claim~C2 - Shani Louk found alive (1 post):} Shani Louk was a German-Israeli civilian abducted during the October~7, 2023 Hamas-led attack in southern Israel. Footage of her circulated globally in the first days of the war. One post in our data amplifies the false claim that she was later ``found alive in a Gaza hospital.'' Her death was confirmed on October~30, 2023, supported by forensic evidence from the Abu Kabir Institute indicating she was killed during or shortly after the October~7 attack~\citep{timesofisrael_louk_2023}. The post received 44~retweet actions from Component~3-SoftWarNews members and 691~total retweets across the full dataset, making it the single most-retweeted text in that component and illustrating how coordinated amplification can rapidly scale a false narrative.

    \item \textbf{Claim~C3 - Al Jazeera staged footage (5 posts):} Allegation that Al Jazeera fabricated wartime visuals by staging ``dead body'' scenes in Gaza. The claim posts present a short clip and argue that movement visible in the video proves the scene was staged. The posts are miscaptioned as the video is from a student protest in Egypt from 2013 and unrelated to the Israel-Hamas war~\citep{snopes_aljazeera_2023}. In Component~8-DFRAC, six closely related posts carrying this claim receive 44~of 86~total retweet actions (51.2\%), including five posts retweeted by all eight component members.
\end{itemize}

\subsection{Tweet Types}

\begin{figure}
    \centering
    \includegraphics[width=\textwidth]{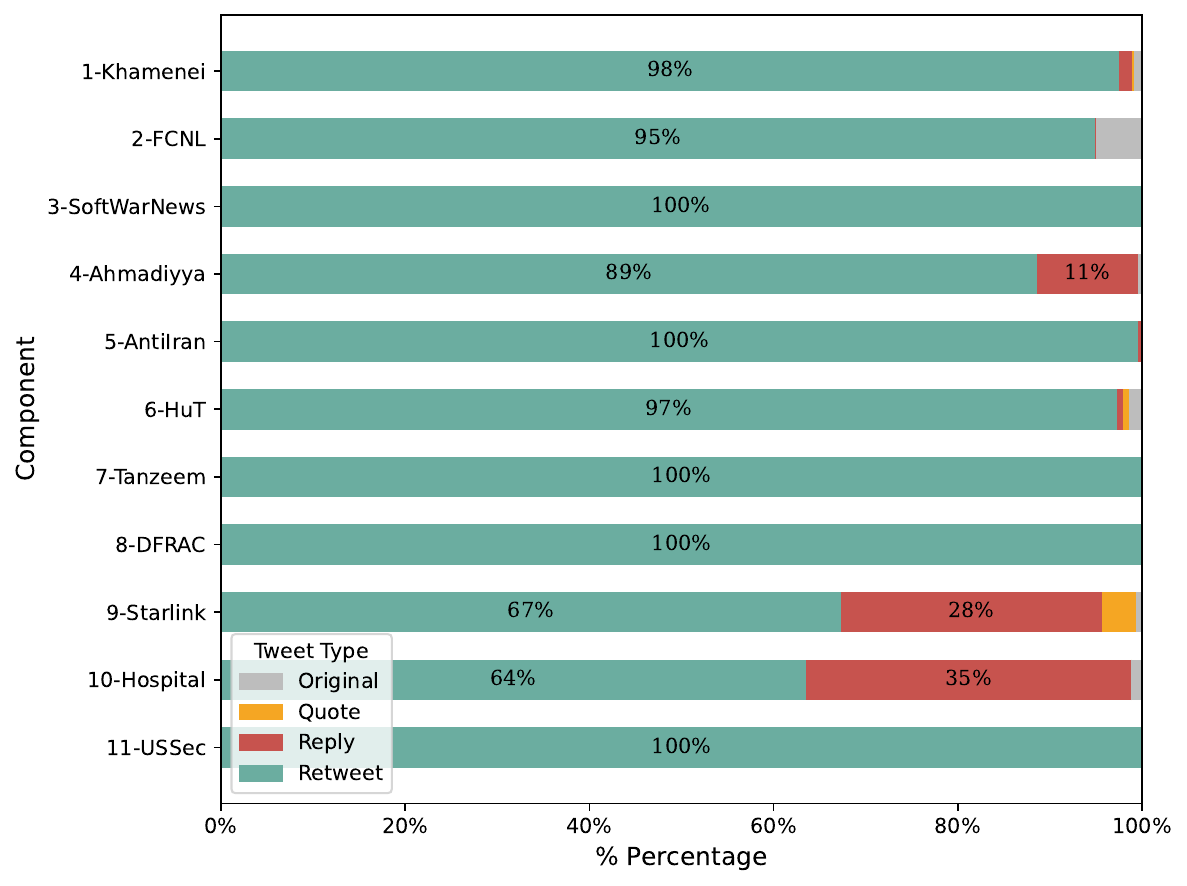}
    \caption{Composition of tweet types in each coordinated component. Bars show the fraction of posts authored by component members that are original tweets, retweets, quoted tweets, or replies.}
    \label{fig:tweet_types}
\end{figure}

To characterize the behavioral patterns of coordinated components, we examine the relative proportions of tweet types (original tweets, retweets, quoted tweets, and replies) within each component. As shown in Figure~\ref{fig:tweet_types}, activity within most components is dominated by retweets, consistent with the overall composition of the dataset (Table~\ref{tab:dataset}). Retweets account for 95\% or more of all activity in eight of the 11 components. The relative scarcity of tweet types typically associated with more organic engagement may suggest that these components are driven primarily by inauthentic or automated activity. However, confirming this interpretation would require additional analysis outside of the current scope.

\subsection{Most Retweeted Accounts}

\begin{figure}
    \centering
    \includegraphics[width=1\textwidth, trim=0.9cm 0 0.28cm 0.5cm, clip]{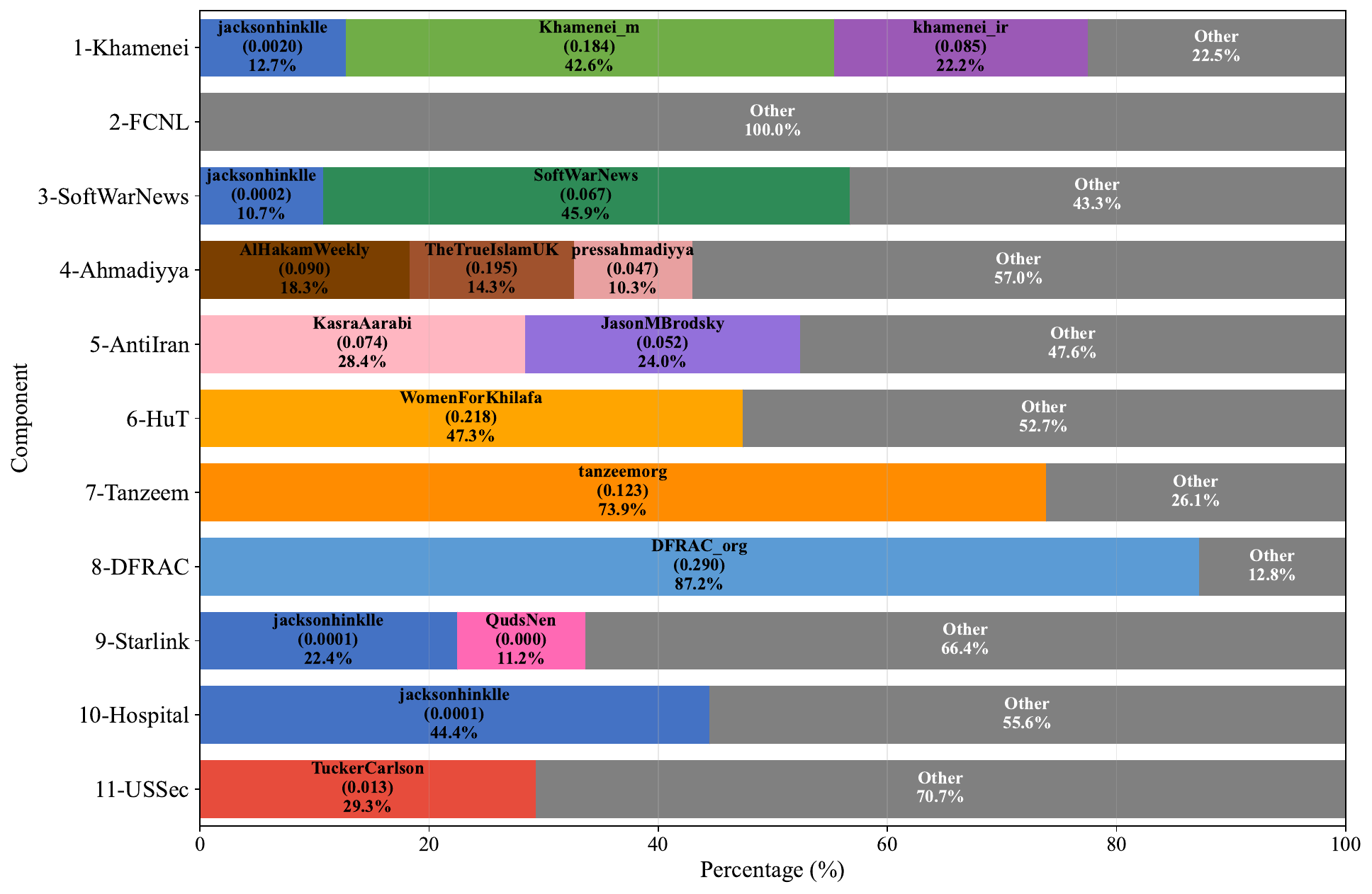}
    \caption{Most frequently retweeted accounts in each coordinated component. For each account with at least 10\% share of retweets, the bars indicate the share of retweets within the component. Coordination reliance is reported in parenthesis. Accounts below 10\% share of retweets are merged into ``Other.''}
    \label{fig:mostretweeted}
\end{figure}

As most components display coordination through retweeting, we analyze the most frequently retweeted accounts (whether or not they are part of the components) to better understand the campaign focus.
For each component, we compute the share of retweets directed to each retweeted account.
Additionally, for each retweeted account, we compute the ratio of retweets they receive from the respective coordinated component to all the retweets they receive. We name this \emph{coordination ncence}.
For each component, Figure~\ref{fig:mostretweeted} reports on the accounts they retweet the most, the proportion of retweets the accounts they receive within the component, and their coordination reliance. To facilitate interpretation, we report only accounts that receive at least 10\% of retweets within a given component. Across most components, retweeting activity is highly concentrated on a small number of accounts that are largely specific to the individual campaigns. The only exception is Jackson Hinkle (@jacksonhinklle), an ideological commentator associated with anti-Zionist narratives, who appears across multiple components. Coordination reliance varies across accounts. Some rely heavily on coordinated groups for retweet amplification, while others such as Jackson Hinkle receive substantial retweeting from a broader audience.

\subsection{Toxicity}

We analyze whether coordinated users express more toxic language than non-coordinated users. We employ the Perspective API~\citep{jigsaw2025perspective} to score each tweet across six dimensions: \textsc{Insult}, \textsc{Threat}, \textsc{Toxicity}, \textsc{Severe Toxicity}, \textsc{Profanity}, and \textsc{Identity Attack}. Scores range from zero to one, where higher values indicate greater confidence in the presence of the attribute. We aggregate to the user level by computing each user's mean score across all their tweets, so that high-volume accounts do not disproportionately influence the comparison.

For each component, we apply a one-sided Mann-Whitney U test comparing the per-user mean scores of component members against the non-coordinated baseline (all users not assigned to any component). This non-parametric test is appropriate because aggregated toxicity scores remain bounded and right-skewed. Since we conduct 66 tests (11 components $\times$ 6 dimensions), we apply layered Bonferroni correction to control the familywise error rate.

\begin{table*}
    \caption{Median user toxicity scores for each coordinated component and the non-coordinated baseline. Effect reports rank-biserial correlation $r_{rb}$ for the \textsc{Toxicity} comparison. Significance markers use layered Bonferroni correction across all 66 one-sided Mann-Whitney U tests (component vs.\ non-coordinated users): $^{***}$~$p < 0.001/66 \approx 1.52\times10^{-5}$; $^{**}$~$p < 0.01/66 \approx 1.52\times10^{-4}$; $^{*}$~$p < 0.05/66 \approx 7.58\times10^{-4}$.}
    \resizebox{\textwidth}{!}{
    \begin{tabular}{l|l|l|l|l|l||l|c}
    \hline
    \centering
     & & & \textsc{Severe} & & \textsc{Identity} & & \\
     Component & \textsc{Insult} & \textsc{Threat} & \textsc{Toxicity} & \textsc{Profanity} & \textsc{Attack} & \textsc{Toxicity} & Effect\\
     \hline
     1\,-\,Khamenei & 0.096\textnormal{***} & 0.145\textnormal{***} & 0.023\textnormal{***} & 0.084\textnormal{***} & 0.238\textnormal{***} & 0.248\textnormal{*} & $+0.102$ \\
     2\,-\,FCNL & 0.068 & 0.154\textnormal{*} & 0.015 & 0.064 & 0.209 & 0.229 & $-0.038$ \\
     3\,-\,SoftWarNews & 0.143\textnormal{***} & 0.198\textnormal{***} & 0.040\textnormal{***} & 0.126\textnormal{***} & 0.288\textnormal{***} & 0.286\textnormal{***} & $+0.351$ \\
     4\,-\,Ahmadiyya & 0.048 & 0.075 & 0.010 & 0.050 & 0.142 & 0.150 & $-0.507$ \\
     5\,-\,AntiIran & 0.062 & 0.090 & 0.010 & 0.050 & 0.136 & 0.174 & $-0.358$ \\
     6\,-\,HuT & 0.143\textnormal{***} & 0.233\textnormal{**} & 0.049\textnormal{***} & 0.114\textnormal{***} & 0.328\textnormal{**} & 0.306\textnormal{*} & $+0.462$ \\
     7\,-\,Tanzeem & 0.067 & 0.133 & 0.020\textnormal{*} & 0.062 & 0.184 & 0.202 & $-0.204$ \\
     8\,-\,DFRAC & 0.048 & 0.132 & 0.013 & 0.078 & 0.150 & 0.199 & $-0.265$ \\
     9\,-\,Starlink & 0.048 & 0.126 & 0.014 & 0.056 & 0.177 & 0.183 & $-0.355$ \\
     10\,-\,Hospital & 0.081 & 0.144 & 0.016 & 0.069 & 0.269 & 0.279 & $+0.198$ \\
     11\,-\,USSec & 0.067 & 0.156 & 0.013 & 0.051 & 0.162 & 0.227 & $-0.060$ \\
     \hline
     \textit{Non-coordinated} & \textit{0.062} & \textit{0.115} & \textit{0.014} & \textit{0.059} & \textit{0.215} & \textit{0.237} & \textit{--} \\
    \hline
    \end{tabular}
    }
    \label{tab:toxicity_comparison}
\end{table*}

Table~\ref{tab:toxicity_comparison} reports the median user score for each component, with significance markers indicating elevation relative to non-coordinated users under the layered correction scheme. To complement significance testing, we also report rank-biserial correlation ($r_{rb}$) for the \textsc{Toxicity} comparison in each component. Following \cite{kerby2014simple}, we compute $r_{rb}$ via the simple-difference form, $r_{rb}=2U/(n_1 n_2)-1$ where $U$ is the Mann–Whitney statistic computed from the mean user toxicity scores, $n_1$ is the number of users in the coordinated component, and $n_2$ is the number of non-coordinated users. This expression is equivalent to $P(X > Y) - P(Y > X)$, where $X$ and $Y$ denote randomly selected user-level toxicity scores from the component and baseline groups, respectively. Unlike a mean-difference metric, $r_{rb}$ captures distribution shifts: it quantifies how often a randomly selected coordinated user has a higher toxicity score than a randomly selected non-coordinated user. Positive values indicate systematic right-shift in the component distribution, while negative values indicate the opposite. This interpretation is useful here because if elevated toxicity were driven only by a few extreme users, $r_{rb}$ would not show a large component-wide shift.

Components 1\,-\,Khamenei, 3\,-\,SoftWarNews, and 6\,-\,HuT show significantly elevated toxicity across all six dimensions, while Component~2\,-\,FCNL shows significant elevation on \textsc{Threat}. For \textsc{Toxicity} effect size, the largest positive shifts are in components 6\,-\,HuT, 3\,-\,SoftWarNews, and 10\,-\,Hospital, even though the latter does not reach Bonferroni-corrected significance for \textsc{Toxicity}.
For components with 6--11 users (5\,-\,AntiIran, 8\,-\,DFRAC, 9\,-\,Starlink, 10\,-\,Hospital, 11\,-\,USSec), non-significant results cannot rule out moderate effects due to low statistical power; absence of significance should not be interpreted as absence of difference. We note that the performance of the Perspective API varies across languages. Components 3\,-\,SoftWarNews (primarily Indonesian), 7\,-\,Tanzeem (Arabic/Urdu), and 8\,-\,DFRAC (Hindi) are predominantly non-English, therefore their toxicity measurements should be interpreted with caution given potential differential measurement errors.

\subsection{Emotions}

While toxicity measures capture explicit harmful language, they do not account for the broader emotional tone of communication. Even when content is not overtly abusive, coordinated accounts may still differ from the broader conversation in the emotions they express and evoke.
We hypothesize that coordinated accounts used different emotions compared to non-coordinated users tweeting about the Gaza conflict. To test this hypothesis, we first extract a set of emotions present in each tweet of any type.
We utilize a multilingual BERT model fine-tuned for emotion~\citep{toshifumi_bert_emotion}, which predicts Ekman's~(\citeyear{ekman1992argument}) six basic emotions (\textsc{Anger}, \textsc{Disgust}, \textsc{Fear}, \textsc{Joy}, \textsc{Sadness}, and \textsc{Surprise}), along with a neutral class. The model accepts the tweet text as input and outputs a probability score for each emotion. Mirroring the toxicity analysis, we aggregate to the user level by computing each user's mean score across all their tweets, compare each component against the non-coordinated baseline using one-sided Mann-Whitney U tests, and apply the same layered Bonferroni correction scheme as in the toxicity analysis.
Table~\ref{tab:emotion_pvalues} reports median user scores with significance markers, along with rank-biserial correlation effect for \textsc{Fear}.
We observe that many results have $p < 0.001$ before Bonferroni correction but are not significant after correction.
This is especially the case with \textsc{Fear}, which is elevated in eight of the 11 components.
We additionally mark the results when the uncorrected $p < 0.001$, even if they do not meet the conservative Bonferroni significance criterion.

\begin{table*}
    \caption{Median user emotion scores for each coordinated component and the non-coordinated baseline. Effect reports rank-biserial correlation $r_{rb}$ for the \textsc{Fear} comparison. Significance markers are reported when the component median exceeds the baseline median, using the same layered Bonferroni correction used in the toxicity analysis. $^{\dagger}$ denotes uncorrected $p < 0.001$ for emotions above the Bonferroni significance threshold.
    }
    \resizebox{\textwidth}{!}{
    \begin{tabular}{l|l|l|l|l|l||l|c}
    \hline
    \centering
     Component & \textsc{Anger} & \textsc{Disgust} & \textsc{Joy} & \textsc{Sadness} & \textsc{Surprise} & \textsc{Fear} & Effect\\
     \hline
     1\,-\,Khamenei   & 0.154 & 0.001 & 0.008\textnormal{***} & 0.029\textnormal{*} & 0.003 & 0.776\textnormal{***} & $+0.355$ \\
     2\,-\,FCNL       & 0.275\textnormal{**} & 0.003\textnormal{***} & 0.005 & 0.049\textnormal{**} & 0.018\textnormal{**} & 0.598 & $-0.040$ \\
     3\,-\,SoftWarNews & 0.091 & 0.001 & 0.011\textnormal{***} & 0.026 & 0.025\textnormal{***} & 0.730\textnormal{$\dagger$} & $+0.257$ \\
     4\,-\,Ahmadiyya  & 0.165 & 0.001 & 0.006 & 0.040 & 0.004 & 0.768 & $+0.295$ \\
     5\,-\,AntiIran   & 0.118 & 0.001 & 0.004 & 0.008 & 0.004 & 0.832\textnormal{*} & $+0.451$ \\
     6\,-\,HuT        & 0.137 & 0.001 & 0.003 & 0.016 & 0.002 & 0.837\textnormal{$\dagger$} & $+0.452$ \\
     7\,-\,Tanzeem    & 0.038 & 0.001 & 0.003 & 0.005 & 0.001 & 0.933\textnormal{**} & $+0.709$ \\
     8\,-\,DFRAC      & 0.076 & 0.001 & 0.004 & 0.017 & 0.013 & 0.825 & $+0.444$ \\
     9\,-\,Starlink   & 0.236 & 0.002 & 0.010 & 0.132\textnormal{*} & 0.018 & 0.564 & $-0.153$ \\
     10\,-\,Hospital  & 0.281 & 0.002 & 0.005 & 0.044 & 0.009 & 0.524 & $-0.087$ \\
     11\,-\,USSec     & 0.132 & 0.002 & 0.004 & 0.047 & 0.018 & 0.718 & $+0.228$ \\
     \hline
     \textit{Non-coordinated} & \textit{0.173} & \textit{0.002} & \textit{0.004} & \textit{0.025} & \textit{0.008} & \textit{0.613} & \textit{--} \\
    \hline
    \end{tabular}
    }
    \label{tab:emotion_pvalues}
\end{table*}

\textsc{Fear} is the most broadly elevated emotion, reaching Bonferroni significance in components 1\,-\,Khamenei ($^{***}$), 7\,-\,Tanzeem ($^{**}$), and 5\,-\,AntiIran ($^{*}$), and falling just above the Bonferroni significance threshold at uncorrected $p<0.001$ ($^{\dagger}$) in components 3\,-\,SoftWarNews and 6\,-\,HuT. Component~2\,-\,FCNL is a notable exception, showing significant elevation in \textsc{Anger}, \textsc{Disgust}, \textsc{Sadness}, and \textsc{Surprise} but not in \textsc{Fear}, consistent with advocacy-driven outrage rather than threat amplification. Component~9\,-\,Starlink shows significantly elevated \textsc{Sadness}, while components 10\,-\,Hospital and 11\,-\,USSec show no significant emotion elevation under any correction. Per-component emotion results are detailed in Section~\ref{sec:component_narratives}. As with toxicity, emotion results for components 3\,-\,SoftWarNews, 7\,-\,Tanzeem, and 8\,-\,DFRAC may be subject to differential measurement error given the multilingual BERT model's uneven coverage of Indonesian, Arabic/Urdu, and Hindi.

\subsection{Image Themes}

\begin{figure}
\begin{center}
\includegraphics[width=\textwidth, trim=0 0 0 1cm, clip]{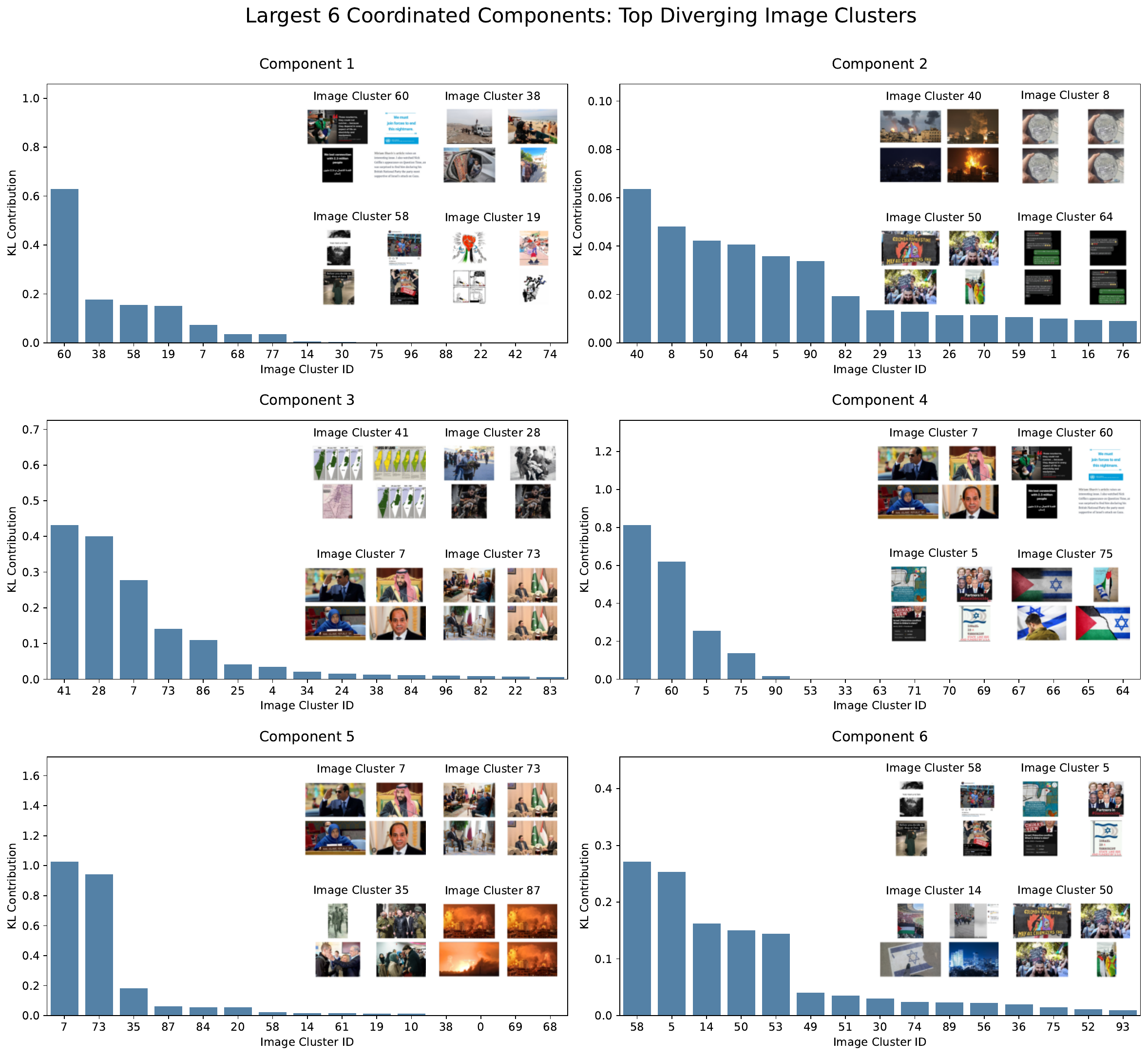}
\caption{Image cluster distributions for the six largest groups of coordinated users.
Images from the four clusters contributing most to the KL divergence between users in each component and all users are shown on the right within each plot.}
\label{fig:top_diverging_img_clusters_by_component}
\end{center}
\end{figure}

We analyze the thematic structure of images shared by coordinated components by comparing their dominant visual themes to those shared by a baseline made by all users in the dataset. To do so, we first cluster images via $K$-means using the BLIP-2 embeddings computed earlier. We set $K=100$, which provides a reasonable within-cluster sum of squares and a relatively uniform number of images per cluster compared to other values ($10 \leq K \leq 200$). This approach provides a coarse but interpretable partition of a large, heterogeneous image corpus. We use the clusters descriptively to compare image-sharing patterns across coordinated components.

For each coordinated component, we compute the empirical distribution of image clusters used by accounts in that component. We do the same for baseline users. To quantify how much a given coordinated component's image usage diverges from baseline behavior, we compute the KL divergence between the component's cluster distribution and the baseline distribution. For each component, we further decompose the KL divergence by image cluster to identify which clusters contribute most to the deviation from the baseline. We treat these clusters as the component's most salient visual themes.

For each of the six largest coordinated components, Figure~\ref{fig:top_diverging_img_clusters_by_component} reports on the image cluster distributions and illustrates the clusters that contribute most significantly to the Kullback-Leibler (KL) divergence between the image-sharing distributions of coordinated and baseline users.
Distinct coordinated components deploy different---though overlapping---visual themes.
Four primary characteristics are prevalent across these components: (i)~political and diplomatic imagery, such as portraits of leaders and official meetings, used to establish legitimacy; (ii)~visuals of conflict, including explosions and destruction, intended to evoke strong emotional responses; (iii)~text-heavy graphics and screenshots, which serve to propagate specific narratives and propaganda; and (iv)~scenes of protest and activism, suggestive of mass mobilization and solidarity.
These findings suggest that coordinated visual communication disproportionately combines emotional appeals, narrative persuasion, and symbolic politics.
Together, these components show how different clusters of coordinated accounts emphasize complementary visual strategies to shape perceptions of the Israel-Hamas War.

\section{Campaign Analysis}
\label{sec:component_narratives}

We now characterize each of the 11 coordinated components in Figure~\ref{fig:components}, by examining their narratives through text and images, assessing their toxicity and emotional content, and highlighting components that amplify misleading, false, or contested claims.

\subsection{Khamenei}

This campaign correspond to the largest connected component in our coordination network. It relies heavily on coordinated retweet amplification and anti-Zionist rhetoric.
Most activity in this component consists of retweets, and the most frequently amplified accounts include two profiles associated with Iran’s Supreme Leader, Ayatollah Ali Khamenei: @khamenei\_ir (``Khamanei International'') and @khamenei\_m (``Khamanei Media'').
This suggests that the coordinated accounts function primarily as an amplification network for Iranian political elites, although confirming whether this activity is state-sponsored would require additional evidence.

The component frequently employs text-based screenshots and political cartoons, often paired with images of destruction, to reinforce narrative.
It exhibits significantly elevated toxicity across all six dimensions
(Table~\ref{tab:toxicity_comparison}),
as well as significantly higher levels of \textsc{Fear},
\textsc{Joy},
and \textsc{Sadness}
(Table~\ref{tab:emotion_pvalues}) compared to non-coordinated users.
We note recurrent high-toxicity posts with delegitimizing formulations, including ``enemy ... evil, malicious, brutal, and bloodthirsty.''
This component is the primary amplifier of Claim~C1 (hospital attribution and ``deleted evidence''; see Section~\ref{sec:characterization} for full annotation and evidence): the hospital-strike variant was retweeted by 143 accounts within the component and the ``deleted evidence'' variant by 99 accounts. The same claim family appears in Component~10\,-\,Hospital, but with lower measured salience.

\subsection{FCNL Ceasefire}

This campaign was detected through co-token coordination, as all the users copy-pasted the same tweet ``In war, civilians always pay the highest price. As the conflict in \#Israel and \#Palestine grows, lawmakers must press for de-escalation and restraint.'' The posts include a link to a campaign ``Urge Congress to call for an immediate ceasefire, de-escalation, and humanitarian access to prevent further civilian harm in Israel and the occupied Palestinian territories.'' The campaign was officially launched by the Friends Committee on National Legislation, the Quakers lobbying organization, and promoted on their website and official Twitter account. Accounts are only coordinated in tweeting this specific text. They have no other original tweet (except for one single tweet in Chinese), but display heavy retweet activity. This suggests that a retweet bot network may have been used for this campaign, though confirmation would require further analysis.

Images shared by this campaign emphasize dramatic conflict scenes and protest imagery to convey urgency and mobilize activism.
No toxicity dimension reaches Bonferroni significance for this component aside from \textsc{Threat},
while emotion analysis shows significantly higher expressions of \textsc{Anger},
\textsc{Disgust},
\textsc{Sadness},
and \textsc{Surprise}
(Table~\ref{tab:emotion_pvalues}).
The isolated \textsc{Threat} elevation is consistent with repeated imminent-harm formulations in representative posts (e.g., ``Israel BOMBED ...'' sequences and warnings that hospitals would be bombed).

\subsection{Indonesian SoftWarNews}

This campaign amplifies tweets in Indonesian language that often highlight the humanitarian impact of bombing. The accounts mostly retweet SoftWarNews, an anonymous news aggregator account tweeting in Indonesian. This account shares war related news without being formally associated with a newspaper or a journalist. They describe themselves ``I stand with the oppressed. The anti-colonial struggle is a humanitarian struggle.'' The account received only 6\% of their retweets from this component, suggesting that they tweet to a wider audience. They have over 120k followers by 2026.

The coordinated component's images focus on maps, territorial claims, and leader portraits, which work to assert geopolitical and legal legitimacy.
This component shows significantly elevated toxicity across all six dimensions
(Table~\ref{tab:toxicity_comparison}), along with significantly higher \textsc{Joy},
\textsc{Surprise},
and \textsc{Fear}
relative to non-coordinated users (Table~\ref{tab:emotion_pvalues}).
The pattern aligns with repeated casualty- and atrocity-centered wording in representative posts (e.g., ``murdered people,'' ``baby killed,'' ``killed in an airstrike'').

The most salient content amplified by this group is Claim~C2, the debunked claim that Shani Louk was ``found alive in a Gaza hospital'' (see Section~\ref{sec:characterization} for full annotation and evidence). Component~3 members retweeted this post 44~times, making it the single most-retweeted text in the component (691~total retweets across the full dataset).

\subsection{Ahmadiyya Peace-Messaging}

This campaign advocates for peace and justice in the conflict region and critiques perceived hypocrisy in global peace discourse. They heavily coordinate in retweeting accounts associated with Ahmadiyya, a Muslim Community based in the UK: @pressahmadiyya (Ahmadiyya's press-office account), @TheTrueIslamUK (Ahmadiyya's UK-facing outreach/education account), @AlHakamWeekly (the account for The Weekly Al Hakam, Ahmadiyya's English-language community publication).
Manual inspection of their other most-retweeted accounts shows that nearly all are associated with the same community, suggesting that this component largely consists of accounts amplifying content from within that community. Almost all non-retweets in this component are replies that are sourced from a single account that replies to prominent pro-Palestinian figures with the same exact text.
The component's images blend diplomatic imagery with national symbols---such as flags---to link leadership with identity-based narratives. No toxicity or emotion dimension reaches Bonferroni significance for this component.

\subsection{Anti-Iran Security-Commentary}

This component promotes a security-focused narrative, linking concerns about radicalization and violence in Europe to developments in the Middle Eastern conflict. Their most retweeted accounts appear to have anti-Iranian regime stance: Jason Brodsky and Kasra Arabi are explicitly affiliated with the non-profit ``United Against Nuclear Iran.''
No toxicity dimension reaches Bonferroni significance for this component. Emotion analysis shows significantly elevated \textsc{Fear}
(Table~\ref{tab:emotion_pvalues}), despite the otherwise neutral linguistic profile.
Visually, this component connects diplomatic events to scenes of conflict, framing political actions as directly tied to security threats.

\subsection{Hizb ut-Tahrir Media-Bias}

This campaign targets perceived Western media bias, particularly that of the BBC, and critiques it as spreading propaganda. They mainly retweet Hizb ut-Tahrir affiliated accounts such as @WomenForKhilafa, @HizbAmerica, @HizbBritain. The account @WomenForKhilafa heavily relies on this group for engagements: the component's retweets make up 21\% of all their retweets.
The component exhibits significantly elevated toxicity across all six dimensions
(Table~\ref{tab:toxicity_comparison}), along with high \textsc{Fear}
(Table~\ref{tab:emotion_pvalues}).
This pattern is supported by representative posts using adversarial formulations such as ``genocidal campaign,'' ``massacre,'' and explicit calls to ``go to war with Israel.''
Visually, it amplifies humanitarian narratives through images depicting destruction, casualties, and protests, emphasizing civilian suffering and rights violations.
Some users frame BBC and Western media through media-bias hashtags and ``propaganda'' narratives, although this is not the most salient pattern in the component.

\subsection{Tanzeem-e-Islami Religious-Solidarity}

This component centers on coordinated retweets of Islamic greetings. They mostly retweet @tanzeemorg, the official account for Tanzeem-e-Islami, a Pakistani Islamic political/ideological organization. They make heavy use of text-centric images, relying on messages aimed at persuasion and expressions of public solidarity. Only \textsc{Severe Toxicity} reaches significance
(Table~\ref{tab:toxicity_comparison}).
The difference in median \textsc{Severe Toxicity} between component users and non-coordinated users is small;
given that this component has only 11 users, this estimate should be interpreted with caution.
The signal appears to be driven by a subset of highly charged posts centered on condemnation of Israeli aggression/``zulm'' in Gaza, \#Gaza\_under\_attack framing, and protest/solidarity appeals, rather than uniformly toxic language across all members. Emotion analysis shows an elevated \textsc{Fear} score
(Table~\ref{tab:emotion_pvalues}), the largest \textsc{Fear} effect across all components, consistent with the alarming/urgent register of solidarity-framed content.

\subsection{DFRAC Fact-Checking}

This campaign amplifies narratives around fact-checking and disinformation in media coverage of the Gaza conflict. They almost exclusively retweet @DFRAC\_org, the official account of the Digital Forensics, Research and Analytics Center, a fact-checking organization based in India. @DFRAC\_org relies on this component for 29\% of their retweets, which may suggest that the component's primary function is to amplify this account. The retweeted posts are primarily in Hindi script; no toxicity or emotion dimension reaches Bonferroni significance for this component. Visually, this component relies heavily on text-centric images, particularly informational screenshots and bureaucratic documents, to reinforce its claims.
The most salient content amplified by this component is Claim~C3, a misleading/decontextualized assertion that Al Jazeera staged ``dead body'' footage in Gaza (see Section~\ref{sec:characterization} for full annotation and evidence). Within this component, the six posts carrying Claim~C3 receive 44 of 86 total retweet actions (51.2\%), including five posts retweeted by all eight component accounts. By contrast, the remaining 42 retweet actions are spread across 30 other posts that include legitimate debunks. This mixture of valid and misleading fact-checking illustrates how the infrastructure of fact-checking can be selectively co-opted for disinformation purposes.

\subsection{Starlink for Gaza}

This campaign was detected through the co-token coordination indicator.
All the accounts in this component copy-paste ``Israel dropped the communication lines for the people in Gaza. They aren't reachable via phone calls or the Internet. They need your help providing them with the internet via Starlink \#starlinkforgaza.''
The campaign's request for Starlink internet access is based on the factual premise of communication blackout in Gaza~\citep{netblocks_gaza_blackout_2023}.
The component accounts tag various pro-Palestinian influencers to receive support from them.
They also retweet a range of pro-Palestinian content, but the amplification is not centered on a small set of prominent or authoritative accounts like the co-retweet components.
No toxicity dimension reaches Bonferroni significance for this component.
Emotion analysis shows significantly elevated \textsc{Sadness}
(Table~\ref{tab:emotion_pvalues}), consistent with the humanitarian urgency and community-solidarity register of the copy-pasta text.
Visually, the component is characterized by a mix of emotionally charged symbols and imagery depicting protests and military actions.

\subsection{Baptist-Hospital Deletion Claim}

This component comprises coordinated accounts that copy-pasted ``@Israel DELETED their fake video claiming to show a Hamas rocket striking the Gaza Baptist Hospital. @HananyaNaftali (Israel's Digital Spokesperson) DELETED his tweet admitting that Israel bombed the Gaza Baptist Hospital. Bad day to be a war propagandist!''
Although the tweets convey strong emotional intensity, no toxicity or emotion dimension reaches Bonferroni significance for this component (Tables~\ref{tab:toxicity_comparison}, \ref{tab:emotion_pvalues}).
The images associated with this campaign are notable for their heavy use of text-based and news-style visuals, including screenshots of headlines, public statements, and posts presented as factual updates.
This component centers on Claim~C1 (hospital attribution and ``deleted evidence''; see Section~\ref{sec:characterization}), operating as a copy-pasta amplifier of the same narrative amplified by Component~1\,-\,Khamenei through retweets.

\subsection{U.S. Right-Wing Security}

This campaign amplifies commentary from right-wing influencers such as Tucker Carlson, republican politicians, and retired U.S. military figure Douglas Macgregor discussing security risks associated with the Gaza-Israel conflict.
No toxicity or emotion dimension reaches Bonferroni significance for this component (Tables~\ref{tab:toxicity_comparison}, \ref{tab:emotion_pvalues}).
Shared images follow a hybrid pattern, combining photos of political elites with text-based infographics to emphasize authority and strategic analysis.
The most amplified post is a Tucker Carlson roundup post that lists a Douglas Macgregor segment and its discussed topics (homeland threats, Iran's missile capacity, Gaza operations, hostages, and economic effects), rather than making a single concrete factual claim. Other recurring posts raise questions about whether U.S.\ weapons reached Hamas from Afghanistan or Ukraine, but these claims appear at lower amplification levels.

\section{Implications for Online Moderation}
\label{sec:risk_synthesis}

Trust-and-safety teams confronting a fragmented coordination landscape must decide where to focus limited moderation resources. This section provides empirical guidance through three analyses based on our case study.
We define \textit{integrity risk} as the likelihood that a coordinated component is actively amplifying misleading claims.
First, from a moderation perspective, intervening on components with high concentrations of misleading claims is more efficient than treating all coordinated activity as equally risky. This efficiency gain, however, only holds if such concentration actually exists. We explore this intervention through a permutation test, confirming that misleading claims cluster in a small subset of components beyond what component size alone would predict.
Second, since misleading claims are rarely known a priori, a natural hypothesis is that elevated toxicity or emotional intensity could serve as reliable proxies to flag high-risk components without costly claim-level annotation. We show this is not the case: misleading-post count, toxicity, and fear elevation are mutually uncorrelated, meaning neither toxic language nor heightened emotion reliably identifies the components spreading false claims, and relying on either would both miss genuine integrity risks and flag components posing none.
Third, we ask how much mitigation would be actually provided by targeted takedown. We show that removing the top amplifiers of known misleading content suppresses a meaningful share of coordinated amplification, though the effect plateaus beyond a small number of accounts. By contrast, removing the top broad amplifiers, a more realistic heuristic when specific misleading posts are unknown, fails to suppress any misleading tweet entirely.
These findings show that effective platform governance requires claim-level misinformation identification: coordination detection alone is necessary but not sufficient.
We now provide each analysis in turn.

\subsection{Integrity-risk Concentration}

We focus on the 191 posts amplified through retweets by at least five accounts in the same coordinated group (Section~\ref{sec:characterization}).
Among these, our annotation identifies 12 misleading posts, which are amplified in only three components.
We test whether such observed concentration could occur by chance under constraints imposed by component size.
Since larger components naturally contain more posts, they have more opportunities to include a misleading post purely by chance.
Therefore, a naive per-component count may conflate genuine concentration with size effects.

To control for this, we conduct a label-permutation test.
Each of the 191 highly retweeted posts is associated with its observed component: for example, 86 posts contributed by Component~1\,-\,Khamenei are attributed to Component~1, and 1 post in Component~11\,-\,USSec is attributed to Component~11.
These component memberships are never changed.
We permute the \textit{integrity labels} of the posts: in each permutation, we draw a new random subset of exactly 12 posts from the full pool of 191 and treat those as the ``misleading'' posts. (The posts selected in different permutations are non-overlapping.) As the draw is uniform over all 191 posts and component sizes differ, larger components are proportionally more likely to receive a misleading label in any given permutation. After each draw, we record how many distinct components contain at least one of the 12 randomly selected posts.
Across 100{,}000 permutations, only 3.9\% produce a distribution equally or more concentrated, i.e., where all 12 randomly drawn posts happen to fall within 3 or fewer components.
This result confirms that the observed degree of concentration is unlikely to arise from component size alone ($p = 0.039$).

\begin{table}
\centering
\caption{Component integrity and behavioral signal profiles. Misleading counts are out of the 191 high-retweet posts (Section~\ref{sec:characterization}). Toxicity columns summarize Bonferroni-significant dimensions and the rank-biserial effect size $r_{rb}$ for \textsc{Toxicity} (Table~\ref{tab:toxicity_comparison}). Emotion columns list Bonferroni-significant emotion dimensions (Fe\,=\,\textsc{Fear}, An\,=\,\textsc{Anger}, Di\,=\,\textsc{Disgust}, Sa\,=\,\textsc{Sadness}, Su\,=\,\textsc{Surprise}, Jo\,=\,\textsc{Joy}; $^{***}p<1.52\times10^{-5}$; $^{**}p<1.52\times10^{-4}$; $^{*}p<7.58\times10^{-4}$; $^{\dagger}$\,uncorrected $p<0.001$; `--' no dimension significant) and the rank-biserial effect size for \textsc{Fear} (Table~\ref{tab:emotion_pvalues}).}
\label{tab:risk_synthesis}
\resizebox{\textwidth}{!}{%
\begin{tabular}{lcp{4.2cm}ccp{3.2cm}c}
\toprule
& & & \textbf{Toxicity} & \textbf{Toxicity} & & \textbf{Fear} \\
\textbf{Component} & \textbf{Misleading} & \textbf{Evidence profile} & \textbf{Signals} & \textbf{$r_{rb}$} & \textbf{Emotion Signals} & \textbf{$r_{rb}$} \\
\midrule
1\,-\,Khamenei     & 6  & Co-retweet amplifier of hospital-attribution claims              & 6/6 & $+0.102$    & Fe***, Jo***, Sa*          & $+0.355$ \\
2\,-\,FCNL         & 0  & Registered lobbying ceasefire campaign                           & 1/6 & $-0.038$ & An**, Di***, Sa**, Su**    & $-0.040$ \\
3\,-\,SoftWarNews  & 1  & Co-retweet amplifier of false ``found alive'' claim              & 6/6 & $+0.351$    & Fe$^{\dagger}$, Jo***, Su***       & $+0.257$ \\
4\,-\,Ahmadiyya    & 0  & Community peace-messaging amplification                          & 0/6 & $-0.507$ & --                         & $+0.295$ \\
5\,-\,AntiIran     & 0  & Security commentary; selective amplification                     & 0/6 & $-0.358$ & Fe*                        & $+0.451$ \\
6\,-\,HuT          & 0  & Media-bias framing; elevated toxicity across all dimensions      & 6/6 & $+0.462$    & Fe$^{\dagger}$             & $+0.452$ \\
7\,-\,Tanzeem      & 0  & Religious solidarity greetings                                   & 1/6 & $-0.204$ & Fe**                       & $+0.709$ \\
8\,-\,DFRAC        & 5  & Co-retweet amplifier of misleading staged-footage claims         & 0/6 & $-0.265$ & --                                       & $+0.444$ \\
9\,-\,Starlink     & 0  & Humanitarian connectivity campaign                               & 0/6 & $-0.355$ & Sa*                        & $-0.153$ \\
10\,-\,Hospital    & 0\textsuperscript{a} & Copy-paste amplifier of hospital ``deleted evidence'' narrative  & 0/6 & $+0.198$    & --  & $-0.087$ \\
11\,-\,USSec       & 0  & Right-wing security commentary; low organic overlap              & 0/6 & $-0.060$ & --                                       & $+0.228$ \\
\bottomrule
\multicolumn{7}{l}{\textsuperscript{a}\footnotesize Component~10 amplifies C1 via copy-paste; its posts did not meet the $\geq 5$-retweet threshold used in annotation.}
\end{tabular}}
\end{table}

\subsection{Correlation between Toxicity, Emotions, and Misleading Claims}

Prior work has found a positive association between hate speech use and low-credibility news exposure~\citep{kim2024toxic}.
An analogous hypothesis on the production side would be that coordinated accounts with elevated toxicity or emotional intensity are also the primary spreaders of misleading claims, and one feature may signal the other. To test this hypothesis, we compute Spearman correlations among the three attributes.
Table~\ref{tab:risk_synthesis} summarizes the profiles of all 11 components.

We find no significant correlation between misleading-post count, \textsc{Toxicity} effect size, and \textsc{Fear} effect size:  Spearman $r_s=0.32$ for misleading\,vs.\,\textsc{Toxicity}, $r_s=-0.05$ for misleading\,vs.\,\textsc{Fear}, and $r_s=-0.03$ for \textsc{Toxicity}\,vs.\,\textsc{Fear}; all $n=11$, $p > 0.3$. Component~8\,-\,DFRAC concentrates five misleading posts yet is one of the least toxic components and less toxic than the non-coordinated users ($r_{rb} = -0.265$), and has no significant emotion elevation; its neutral linguistic framing would cause toxicity or emotion filters to miss the integrity risk entirely. The mirror case is Component~6\,-\,HuT: the highest toxicity effect ($r_{rb} = 0.462$, 6/6 dimensions significant) and elevated \textsc{Fear} ($r_{rb} = 0.452$) with zero misleading posts, showing that inflammatory framing does not imply false claims. Component~7\,-\,Tanzeem similarly produces the dataset's largest \textsc{Fear} elevation ($r_{rb} = 0.709$) with negligible toxicity and no misleading content. The lack of correlation indicates that no single content-based signal is a reliable proxy for the others, and that integrity-risk assessment requires the identification of misleading claims.

\subsection{Targeted Takedown}

Given that the integrity risk is concentrated in a small subset of components, we next ask how much mitigation would be provided by a targeted account takedown. Following \citet{deverna2024identifying}, we carry out a dismantling analysis to evaluate how information integrity is affected by removing the most active amplifiers of misleading claims.

We define a \emph{misleading retweet action} as a single (account, tweet) pair in which a coordinated account retweeted one of the 12 misleading posts. Across the dataset, 178 distinct coordinated accounts performed at least one such action, producing 237 misleading retweet actions in total. Of these 178 accounts, 31 retweeted more than one misleading tweet: specifically, eight accounts retweeted five misleading tweets each, one retweeted four, two retweeted three, 20 retweeted two, and 147 retweeted exactly one.

We compare two experimental setups that differ in the information assumed to be available to the intervening party, such as a platform's content moderation team. This affects the order in which accounts are removed:
\begin{enumerate}
    \item \textbf{Known misleading tweets:} We assume the 12 misleading tweets are known a priori. We rank all 178 accounts by their total number of misleading retweet actions and simulate removing the top-$k$ accounts for increasing $k$. This represents an idealized scenario in which content moderation has already identified which specific posts carry false or misleading claims.
    \item \textbf{Heuristic (no prior knowledge):} In practice, a platform may not know in advance which posts are misleading. As a baseline, we consider a heuristic that treats all 191 widely-amplified posts as potentially misleading and target the most active amplifiers of this broader pool. We therefore rank all 532 accounts that retweeted any of the 191 widely amplified posts by their total number of retweets across this set, and simulate removing the top-$k$.
\end{enumerate}

For both setups, we track two outcomes as accounts are progressively removed:
\begin{itemize}
    \item \emph{removed actions:} the share of the 237 misleading retweet actions attributed to the removed accounts; and
    \item \emph{fully suppressed tweets:} misleading tweets for which \emph{every} coordinated retweeter falls within the removed set, eliminating all coordinated amplification.
\end{itemize}

\begin{figure}
\centering
\includegraphics[width=\textwidth, trim=0 0 0 1.2cm, clip]{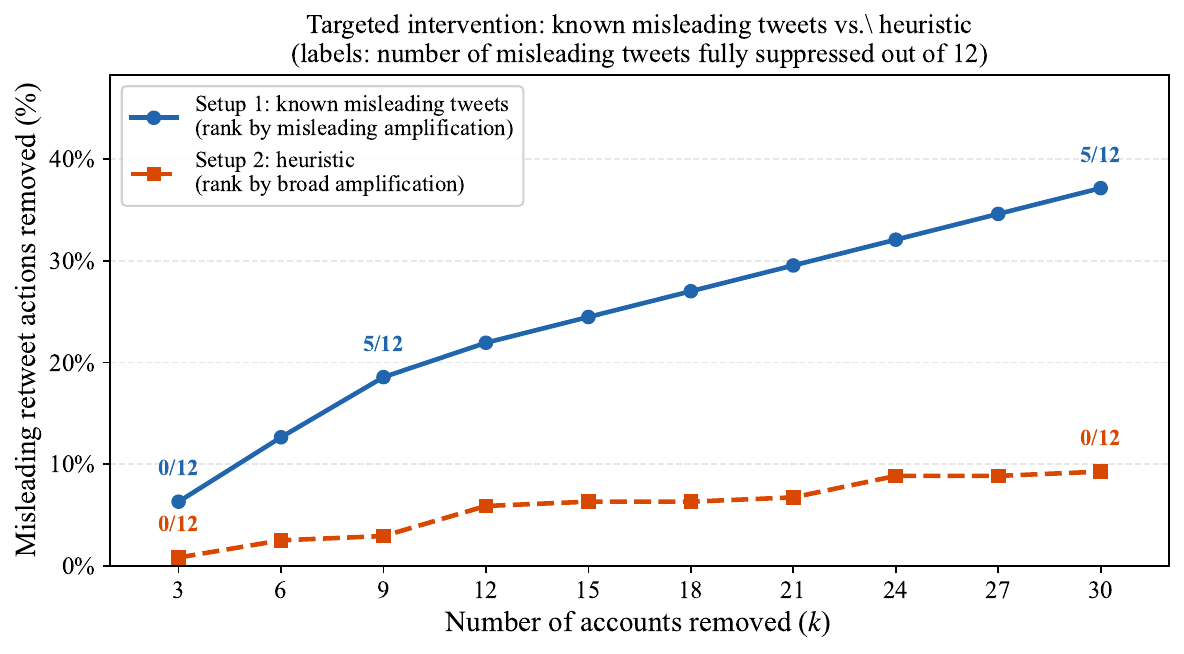}
\caption{Top-$k$ amplifier removal comparing two experimental setups. The $y$-axis shows the percentage of the 237 baseline misleading retweet actions removed. Annotations show the number of misleading tweets fully suppressed at each change point.
}
\label{fig:topk_intervention}
\end{figure}

Figure~\ref{fig:topk_intervention} compares the two setups for $3 \leq k \leq 30$.
Setup~1 is substantially more targeted: at $k = 9$ (5\% of all coordinated accounts with a retweet action), 18.6\% of misleading retweet actions are already removed and 5 of 12 misleading tweets are fully suppressed; by $k = 30$ (17\%), 37.1\% of actions are removed with 5 of 12 suppressed.
Setup~2 removes far fewer misleading actions at every $k$ and suppresses none of the misleading tweets throughout. This is because the the baseline heuristic removes accounts that are heavy amplifiers of advocacy content rather than the specific retweeters of the misleading claims. The contrast illustrates that targeted takedown requires prior identification of the misleading posts.

\section{Discussion}

We study coordinated groups on Twitter during the onset of the 2023 Israel-Hamas conflict. We use multiple indicators to detect 11 components in the coordination network. Analysis of these components, consisting of 541 accounts, tells us that coordination during this high-salience conflict did not manifest as a single, orchestrated campaign but as a structurally fragmented ecosystem of parallel groups.

The components differ in coordination mechanisms, retweet target sets, and content signatures. Each operates in parallel, pursuing its own narrative agenda (Figure~\ref{fig:mostretweeted}). These observations extend to the Israel-Hamas information environment the heterogeneity documented in prior conflict-coordination studies~\citep{alieva2024propaganda,zia2023analysis,luceri2024unmasking}. As a practical consequence, the coordinated ecosystem has no single hub whose removal would neutralize the whole.
Interventions must be designed and applied at the component level rather than treating all coordinating groups as a monolithic campaign.
Counter-messaging or prebunking efforts and other interventions can then be tailored to the narrative and target audience of each component separately, a response more likely to produce results compared to broad-spectrum reaction.

The information-integrity risk is not uniformly distributed across coordinated groups: misleading, false, or contested claims concentrate in a small subset of components.
The component-specificity of integrity risk has a direct implication for moderation design: detecting coordination is a necessary first step, but it is not a reliable guide to where harm is occurring.
Most coordinated behavior in this ecosystem covers advocacy, solidarity, or humanitarian organizing, so treating coordination as a proxy for harm would misclassify the majority of components and risk suppressing legitimate collective action.
At the same time, the highest-risk actors do not announce themselves through elevated toxicity or emotion signals. Component~8's misleading claims are wrapped in a credibility-enhancing fact-checking frame that reduces their visibility to automated filters.
Component-level content profiling for information integrity is therefore a necessary complement to coordination detection.

A persistent challenge in coordination research is distinguishing coordinated inauthentic behavior from legitimate collective action~\citep{nizzoli2021coordinated}, and the coordination signals analyzed here cannot resolve this distinction definitively. Many components such as components 2\,-\,FCNL, 4\,-\,Ahmadiyya, 7\,-\,Tanzeem, and 9\,-\,Starlink display coordination patterns consistent with open, attributable organizing. Their messages link openly to identified organizations or campaigns, and the accounts they amplify do not depend on the coordinated component for the bulk of their retweets. Their low organic engagement is consistent with narrow-audience mobilization. Without account-level behavioral data, however, this interpretation cannot be confirmed.
By contrast, components 1\,-\,Khamenei, 3\,-\,SoftWarNews, 6\,-\,HuT, and 8\,-\,DFRAC exhibit patterns more suggestive of coordinated inauthentic activity --- heavy retweet concentration on narrow targets, near-zero original content, copy-paste templates, and in Component~8 a fact-checking fa\c{c}ade that mixes valid debunks with misleading claims. Yet, definitive attribution would require additional account-level evidence. This continuum underscores why coordination detection alone is an insufficient basis for moderation: the same structural signal (coordinated retweeting) manifests across the full range from plausibly open civic organizing to patterns suggestive of influence operations.

For platform governance, the finding that claim-level integrity, toxicity, and emotional tone are mutually uncorrelated across components is also significant. Each signal captures real variation, but is not a proxy for the others.
The starkest illustration is the  contrast between DFRAC and HuT campaigns: DFRAC concentrates five misleading posts yet shows the most negative toxicity effect in the dataset and no significant emotion elevation, so its fact-checking fa\c{c}ade would evade both toxicity and emotion-based filters entirely; HuT shows the strongest toxicity signals and significantly elevated fear emotion, yet zero misleading posts, demonstrating that inflammatory framing does not imply false claims. Each dimension may capture the information the others may miss, and thus, no single signal is individually sufficient for characterizing component-level content risk.

A key practical implication of our analysis for misinformation governance is \emph{prioritization}. Interventions such as account suspensions or de-ranking should be concentrated on the components where coordinated behavior and integrity-risk evidence co-occur, while components showing coordination without detected integrity risk should be treated as potentially legitimate collective mobilization pending further evidence. This framing provides a practical bridge between network science (component structure) and content moderation (content integrity signals), grounding triage decisions in convergent evidence rather than any single behavioral proxy.

Several limitations qualify these findings.
The analysis covers a single platform, Twitter, during a specific temporal window that may affect the composition of detected components.
We also focus on a limited set of coordination signals, apply strong thresholds to edges in the coordination network, and exclude small connected components.
The impact of all these filters on the composition of detected components cannot be fully assessed, as they may have obscured coordination activity operating through excluded content; this is a known limitation of working with platform-filtered datasets and unsupervised detection methods.

The statistical power of toxicity and emotion tests is limited by group size; for the smallest components, non-significant results cannot rule out moderate effects. The strict Bonferroni correction makes our significant findings conservative in order to reduce the risk of false positives.
In addition, both the Perspective API (toxicity) and the multilingual BERT model (emotions) have uneven cross-language performance; results for the predominantly non-English components, namely 3\,-\,SoftWarNews (Indonesian), 7\,-\,Tanzeem (Arabic/Urdu), and 8\,-\,DFRAC (Hindi), should be interpreted with caution given potential differential measurement error.

Our takedown analysis models the coordination network as static and does not account for replacement accounts, behavioral adaptation, or cross-platform spillover. Additionally, it captures only misleading retweet actions, leaving copy-paste components (2, 9, and 10) entirely outside of its scope.
The absolute counts are small, so the reported percentage reductions should be read as illustrative of the relative effect of targeted enforcement rather than as precise real-world estimates.

Future work should validate whether the same structural archetypes, risk concentrations, and signal decoupling recur in other conflicts and on other platforms, such as Telegram, TikTok, and YouTube; measure cross-platform spillover between components identified here; and explore causal mechanisms linking coordination structure to content integrity outcomes, which the present correlational analysis cannot establish.

\section*{Ethics Statement}

This work analyzes publicly available social media data (tweets) collected under Twitter's Terms of Service. The dataset consists exclusively of public posts; no private data, direct messages, or account credentials were accessed.
The study has been granted exemption from Institutional Review Board review (Indiana University Bloomington protocols 12410 and 1102004860).

\section{Acknowledgements}

ChatGPT was used to support the editing of this manuscript. All content was critically reviewed, verified, and revised by the authors, who take full responsibility for the final manuscript.

This work  was  supported  in part  by DARPA (contract HR001121C0169), the Knight Foundation, and the Lilly Endowment, Inc. through its support for the Indiana University Bloomington Pervasive Technology Institute.

\bibliography{references}

\end{document}